\documentclass[fleqn,12pt,twoside]{article}
\usepackage{espcrc1}

\usepackage{graphics}
\usepackage{epsfig}

\newcommand{\AmS}{{\protect\the\textfont2
  A\kern-.1667em\lower.5ex\hbox{M}\kern-.125emS}}

\def\Journal#1#2#3#4{{#1}{#2} (#4) #3 }

\def\IJMPA{{Int. J. Mod. Phys.}~{A}\,}

\def\NPB{{Nucl. Phys.}~{B}\,}
\def\PLB{{Phys. Lett.}~{B}\,}
\def\PLC{Phys. Repts.\ }
\def\PRL{Phys. Rev. Lett.\ }
\def\PR{Phys. Rev.\ }
\def\PRD{{Phys. Rev.}~{D}\,}

\def\ARNS{{Ann. Rev. Nucl. Part. Sci.\ }} 
\def\RMP{Rev. Mod. Phys.\ }

\title{From Bjorken Scaling to pQCD---Experimental techniques from p-p collisions of the 1970's with application to Au+Au collisions at RHIC.}

\author{M. J. Tannenbaum\address{Physics Dept., 510c\\
 Brookhaven National Laboratory \\ 
        Upton, NY 11973-5000 USA}%
        \thanks{Research supported by U.S. Department of Energy, DE-AC02-98CH10886.}}       
\begin{document}

\maketitle

\begin{abstract}
Hard scattering in p-p collisions was discovered at the CERN ISR in 1972, by the
method of leading particles, which proved that the partons of Deeply Inelastic
Scattering strongly interacted with each other. Further ISR measurements utilizing
inclusive single or pairs of hadrons established that high $p_T$ particles are
produced from states with two roughly back-to-back jets which are the result of
scattering of constituents of the nucleons as described by Quantum
Chromodynamics. These techniques,
which are the only practical method to study hard-scattering and jet phenomena in
Au+Au collisions at RHIC, will be reviewed. 
\end{abstract}

\section{Introduction}

   It is not generally realized that hard-scattering was discovered---in both Deeply Inelastic lepton scattering (DIS) at SLAC~\cite{DIS} and in high $p_T$ particle production in p-p collisions~\cite{CCR,SS,BS} at the CERN ISR---before the discovery of QCD. In that  era, scaling laws relating the transverse momentum ($p_T$) spectra at different values of center-of-mass (c.m.) energy, $\sqrt{s}$, were the key to understanding the underlying physics. Absolute cross sections played a minimal role. Measurements of high $p_T$ particle production near mid-rapidity, at symmetric p-p or A+A colliders, are ideal for such scaling studies---most systematic errors from varying the $\sqrt{s}$ or the species cancel since the detectors are at rest near $90^{\circ}$ in the c.m. system of the reaction and are thus insensitive to longitudinal effects. 
   
 \section{Bjorken scaling and the parton model}
 	   The idea of hard-scattering in N-N collisions dates from the first indication of pointlike structure inside the proton, in 1968, found by deeply inelastic electron-proton scattering~\cite{DIS}, i.e. scattering with large values of 4-momentum transfer squared, $Q^2$, and energy loss, $\nu$. The discovery that the Deeply Inelastic Scattering (DIS) structure function 
\begin{equation}
F_2(Q^2, \nu)=F_2({Q^2 \over \nu})
\label{eq:F2scales}
\end{equation}
 ``scaled'' i.e just depended on the ratio 
\begin{equation}
x=\frac{Q^2}{2M\nu}
\label{eq:x_def}
\end{equation} 
independently of $Q^2$, as originally suggested by 
Bjorken~\cite{Bj}, led to the concept of a proton 
composed of point-like `partons'. The deeply inelastic scattering of an electron from a proton is simply quasi-elastic scattering of the electron from point-like partons of effective mass $Mx$, with quasi-elastic energy loss, $\nu=Q^2/2Mx$. The probability for a parton to carry 
a fraction $x$ of the proton's momentum is measured by $F_2(x)/x$. 

	Since the partons of DIS are charged, and hence must scatter electromagnetically from each other, Berman, Bjorken and Kogut (BBK)~\cite{BBK} and subsequent authors~\cite{CIM,CGKS}, derived a general formula for the cross section of the inclusive reaction  
\begin{equation}
 p + p\rightarrow C +X 
\label{eq:bbk1}
\end{equation} 
using the principle of factorization of the reaction into parton distribution functions for the protons, fragmentation functions to particle $C$ for the scattered partons and a short-distance parton-parton hard scattering cross section. The invariant cross section for the inclusive reaction (Eq.~\ref{eq:bbk1}), where particle $C$ has transverse momentum $p_T$ near mid-rapidity was given by the general `scaling' form:~\cite{CIM}
\begin{equation}
E \frac{d^3\sigma}{dp^3}=\frac{1}{p_T^{n}} F({2 p_T \over \sqrt{s}})
= \frac{1}{\sqrt{s}^{\,n}} G({x_T})\quad \mbox{where}\quad x_T=2p_T/\sqrt{s} \qquad .
\label{eq:bbg}
\end{equation}
The cross section has 2 factors, a function $F$ $(G)$ which `scales', i.e. depends only on the ratio of momenta; and a dimensioned factor, ${p_T^{-n}}$   $(\sqrt{s}^{\,-n})$,   
where $n$ gives the form of the force-law 
between constituents. For QED or Vector Gluon exchange~\cite{BBK}, $n=4$, and for the case of quark-meson scattering by the exchange of a quark~\cite{CIM}, $n$=8. When QCD is added to the mix~\cite{CGKS}, pure scaling breaks down and $n$ varies according to the $x_T$ and $\sqrt{s}$ regions used in the comparison, $n\rightarrow n(x_T, \sqrt{s})$. 

\section{ISR data, notably CCR, CCRS, CCOR, 1972-1978} 
\begin{figure}[ht]
\begin{center}
\begin{tabular}{cc}
\includegraphics[width=0.45\linewidth]{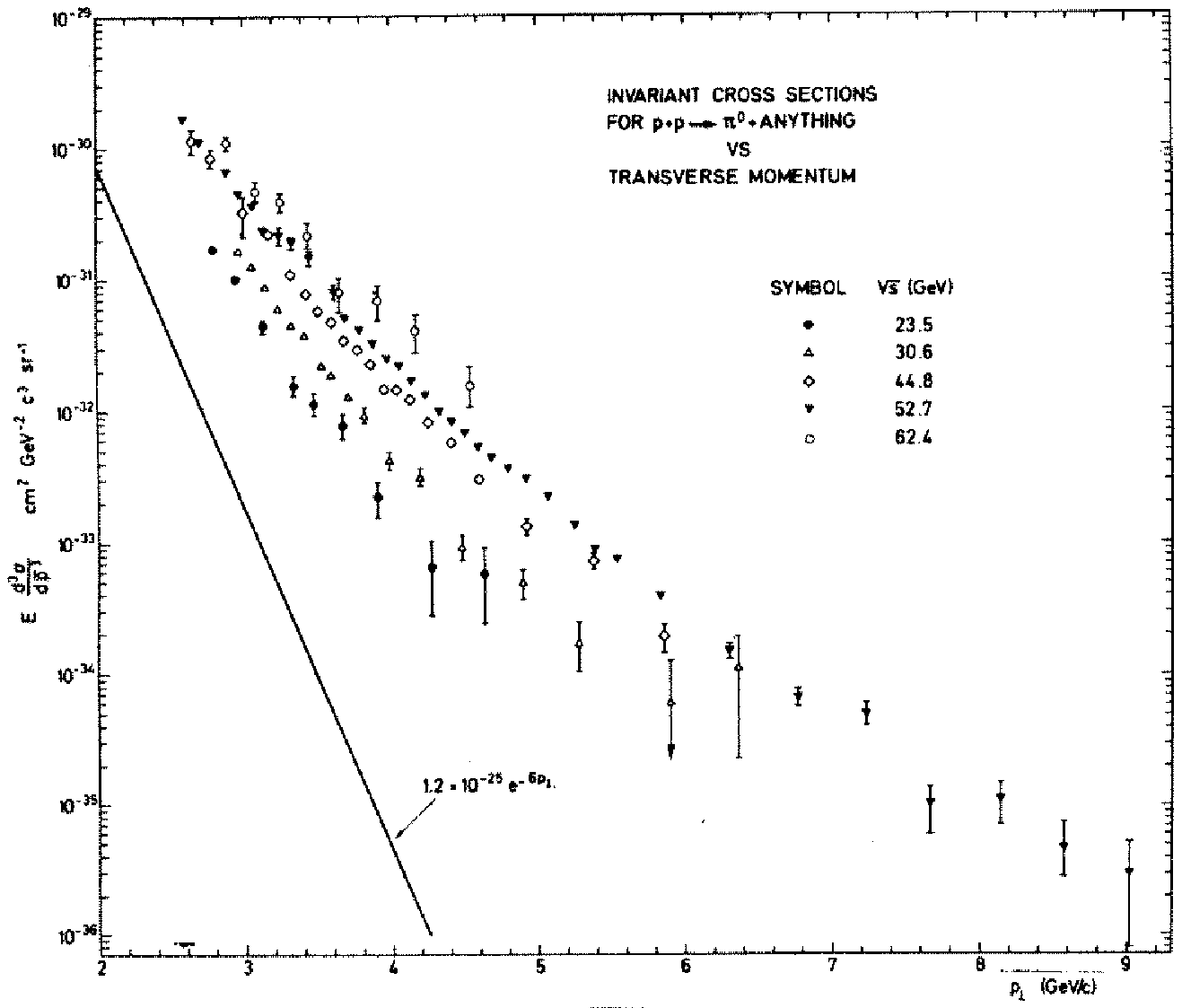} &
\includegraphics[width=0.52\linewidth]{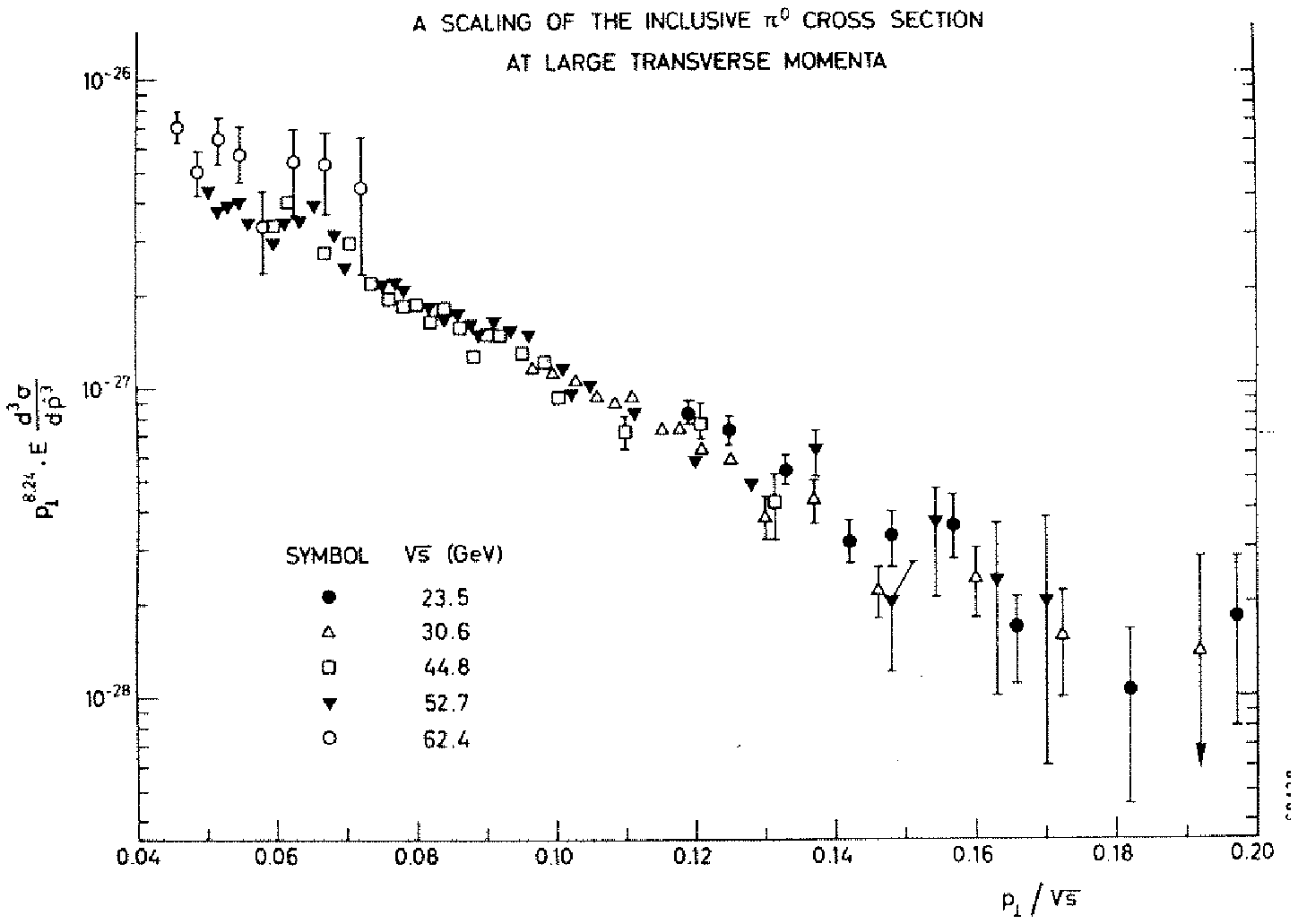}
\end{tabular}
\end{center}
\vspace*{-10mm}
\caption[]
{a) (left) CCR~\cite{CCR} transverse momentum dependence of the invariant cross section at five center of mass energies. 
b) (right) The same data multiplied by $p_{\perp}^n$, using the best fit 
value of $n=8.24\pm0.05$, with $F=Ae^{-b x_{\perp}}$, plotted vs 
$p_{\perp}/\sqrt{s}$.   
\label{fig:CCR} }
\end{figure}

   It was known, from cosmic ray physics~\cite{Cocconi}, that the average transverse momentum of secondary particles in N-N collisions was limited to ~0.5 GeV/c, independent of the primary energy, and could be described by the simple `thermal' form: 
   \begin{equation}
{d\sigma \over {p_T dp_T}}=A e^{-6p_T} \qquad , 
\label{eq:CKP}
\end{equation}
where $p_T$ is the transverse momentum in GeV/c and 
$\langle p_T\rangle=2/6=$0.333 GeV/c. 
The CERN Columbia Rockefeller (CCR) Collaboration~\cite{CCR} 
(and also the Saclay Strasbourg~\cite{SS} and British Scandinavian~\cite{BS} 
collaborations) at the CERN-ISR measured 
pion production over a large range of transverse momenta, unavailable in cosmic ray studies or at lower $\sqrt{s}$. 
The $e^{-6p_T}$ breaks to a power law at high $p_T$ with 
characteristic $\sqrt{s}$ dependence (Fig.~\ref{fig:CCR}). 
The large rate indicates that {\em partons interact 
strongly ($\gg$ EM) with each other}.  
However, the electromagnetic form of BBK~\cite{BBK}, $p_{\perp}^{-4} F(p_{\perp}/\sqrt{s})$, was 
not observed in the experiment. On the other hand, the constituent exchange 
model~\cite{CIM} seemed to give an excellent account of the data~\cite{CCR}.   
The data fit 
$p_{\perp}^{-n} F(p_{\perp}/\sqrt{s})$, with $n\simeq8$. 

The best data at FNAL in 1977~\cite{Cronin} also beautifully showed the CIM scaling with $n\simeq 8$ over the range $0.2\leq x_T\leq 0.6$, for 200, 300 and 400 GeV incident energies. However, this effect turned out not to be due to CIM, but to  the `broadening' by initial state transverse momentum, the ``$k_T$ effect".   

\section{Experimental improvements, theoretical improvements} 
\begin{figure}[ht]
\begin{center}
\begin{tabular}{cc}
\psfig{file=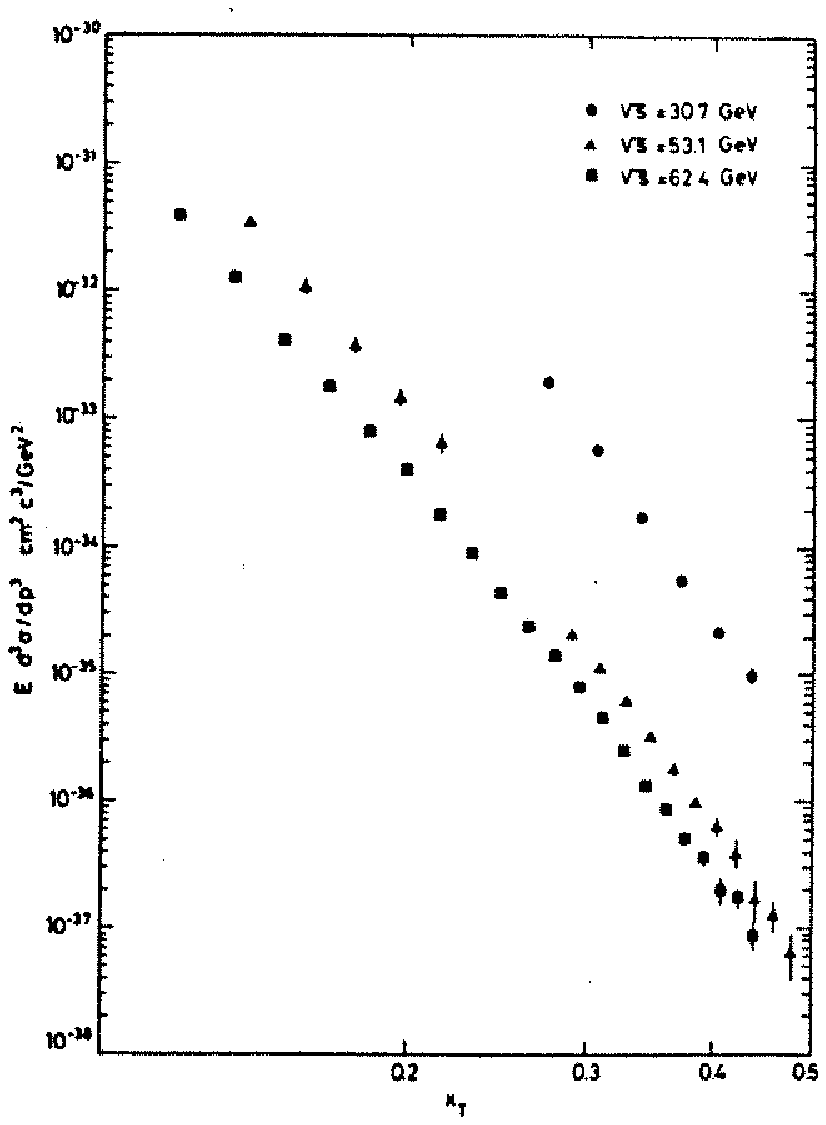,height=3.85in}
\hspace*{0.55in}
\psfig{file=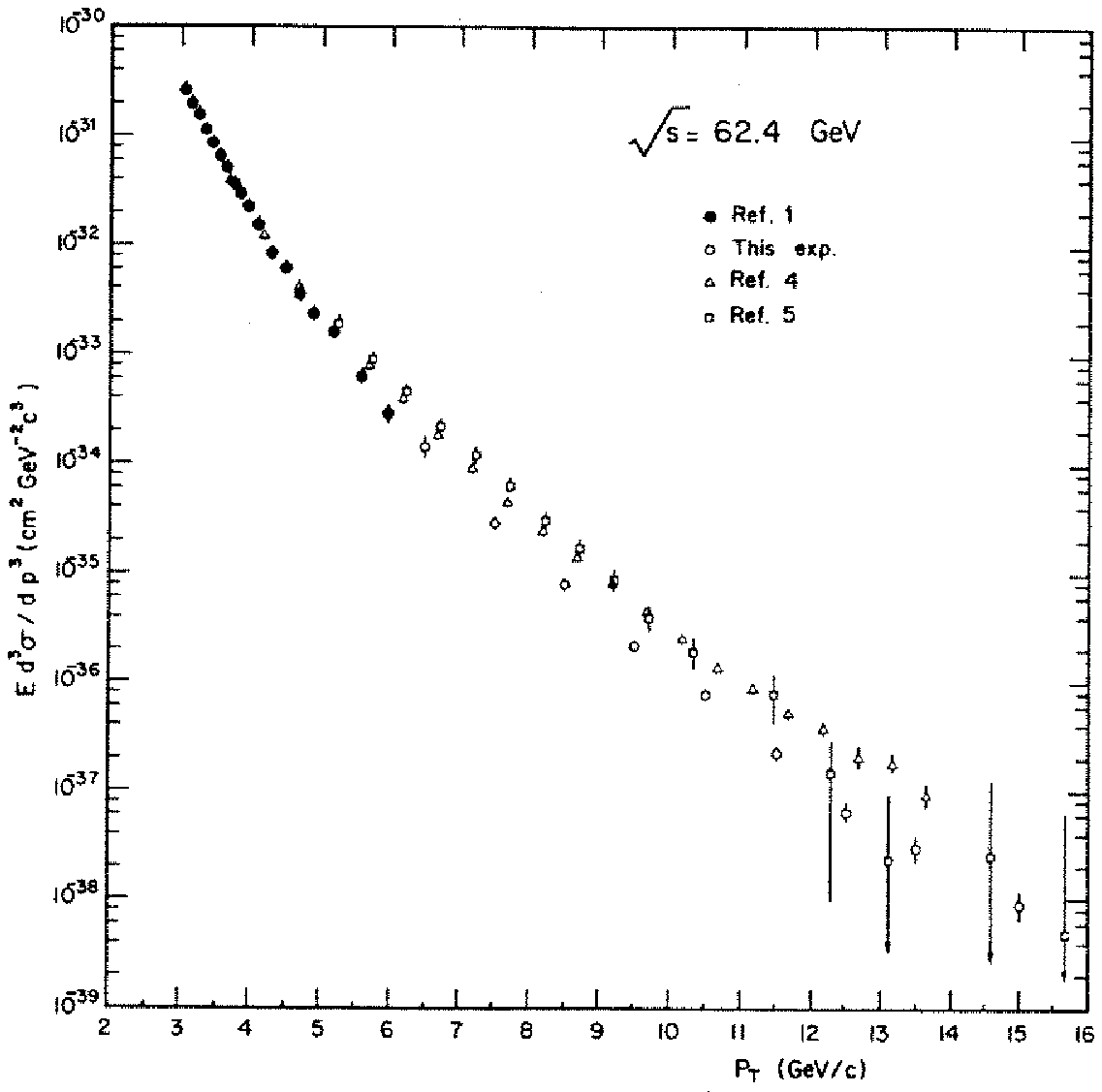,height=3.85in,width=2.75in}\\
\hspace*{0.25in}
\psfig{file=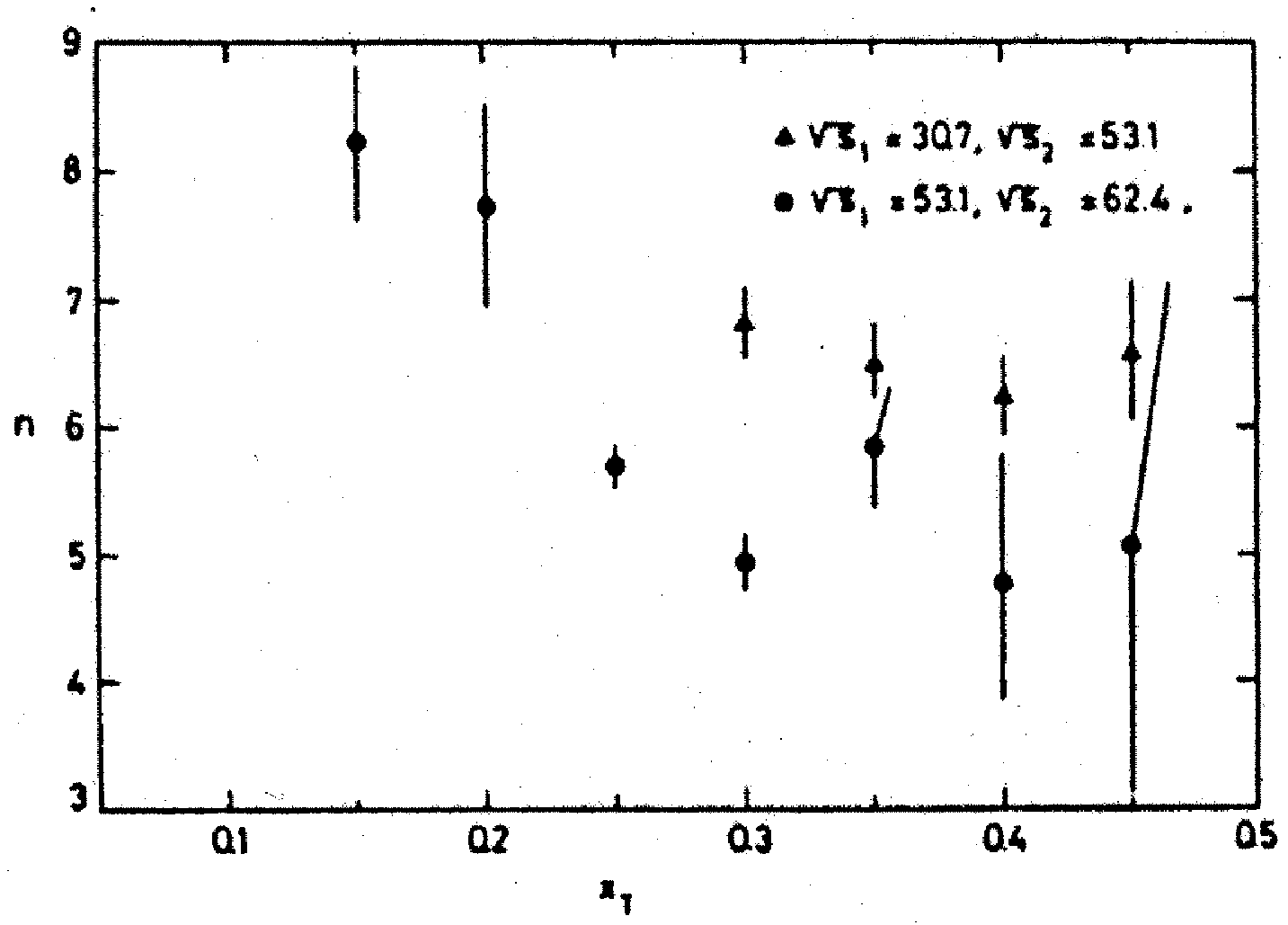,height=1.65in}\hspace*{0.80in}
\psfig{file=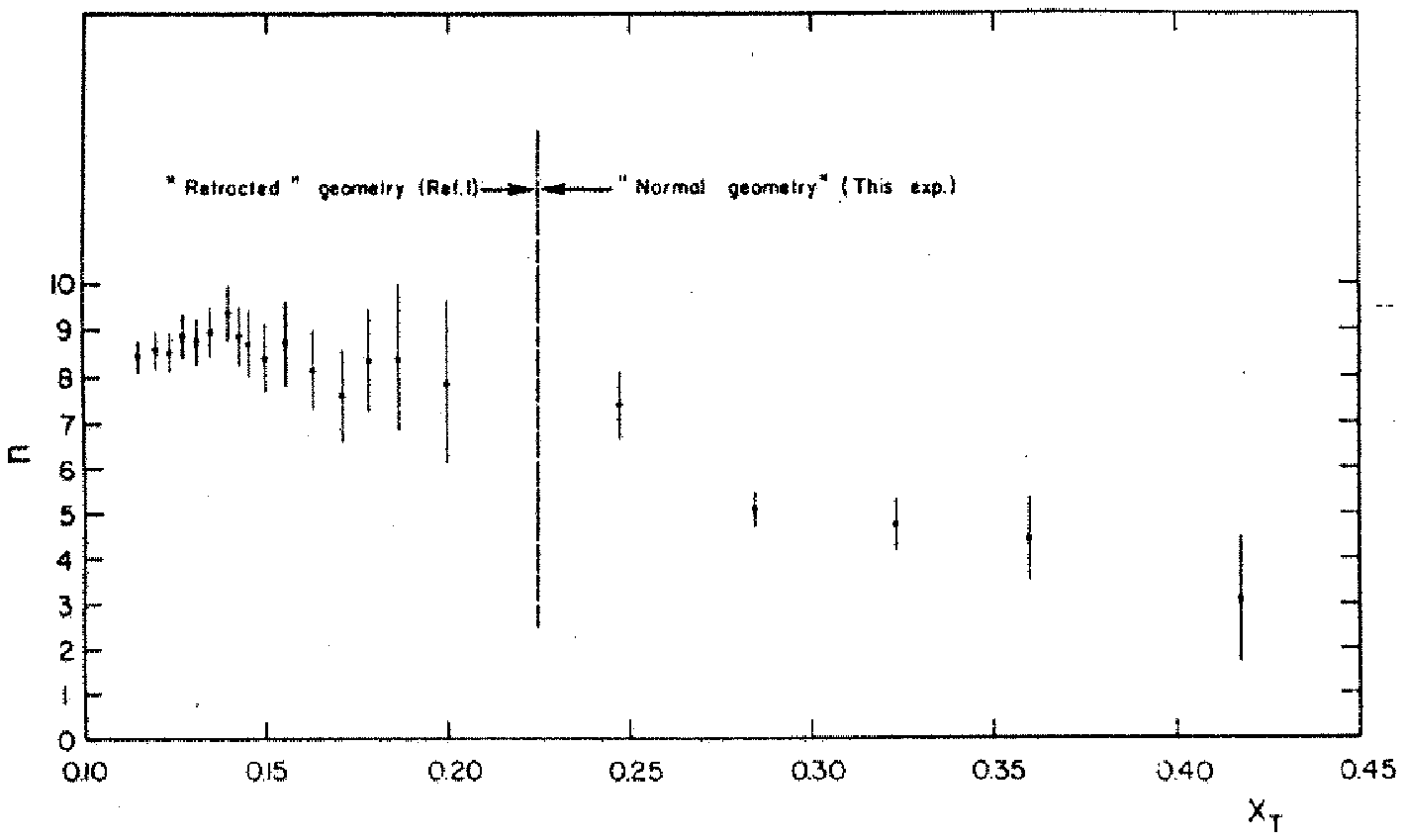,height=1.65in}
\end{tabular}
\end{center}
\vspace*{-14.5mm}
\caption[]
{a) (top-left) Log-log plot of CCOR invariant cross section vs $x_T=2 p_T/\sqrt{s}$; 
b)~(bottom-left) $n(x_T,\sqrt{s})$ derived from the combinations indicated.  
There is an additional common systematic error of 
$\pm 0.33$ in $n$. 
c) (top-right)  Invariant cross section for $\pi^0$ inclusive for several ISR 
experiments, compiled by ABCS Collaboration. d) (bottom-right) $n(x_T,\sqrt{s})$ 
from ABCS 52.7, 62.4 data only. There is an 
additional common systematic error of $\pm0.7$ in $n$.
\label{fig:ccorxt} }
\end{figure}

The CCOR measurement~\cite{CCOR} (Fig.~\ref{fig:ccorxt}a) with a larger apparatus and much increased integrated 
luminosity extended the previous $\pi^0$ measurements~\cite{CCR,CCRS} to 
much higher $p_T$. 
In Fig.~\ref{fig:ccorxt}a, the CCOR $\pi^0$ data for 3 values of $\sqrt{s}$ are plotted vs $x_T$ on a log-log scale. $n(x_T, \sqrt{s})$ is determined for any 2 values of $\sqrt{s}$ by taking 
the ratio of invariant cross sections at fixed $x_T$, with results shown in 
Fig.~\ref{fig:ccorxt}b: $n(x_T, \sqrt{s})$  clearly varies 
with both $\sqrt{s}$ and $x_T$, it is not a constant. For 
$\sqrt{s}=53.1$ and 62.4 GeV, $n(x_T, \sqrt{s})$ varies from $\sim 8$ at low 
$x_T$ to $\sim 5$ at high $x_T$. The new fit~\cite{CCOR}, for $7.5\leq p_T\leq 14.0$ GeV/c, is  
\mbox{$E d^3 \sigma/dp^3\simeq p_T^{-{5.1\pm 0.4}} (1-x_T)^{12.1\pm 0.6}$},  
$53.1\leq \sqrt{s}\leq 62.4$ GeV (including {\em all} systematic errors).

	An important feature of the scaling analysis (Eq.~\ref{eq:bbg}) in  
determining $n(x_T, \sqrt{s})$---{\em is that the absolute $p_T$ 
scale uncertainty and many efficiency and acceptance errors cancel!} The effect of the absoulte scale uncertainty, which 
is the main systematic error in these experiments,  can be gauged from 
Fig.~\ref{fig:ccorxt}c~\cite{ABCS} which shows the $\pi^0$ cross 
sections from several experiments. The absolute cross sections disagree by 
factors of $\sim 3$ for different experiments but the values of 
$n(x_T, \sqrt{s})$ for the CCOR~\cite{CCOR} 
(Fig.~\ref{fig:ccorxt}b) and ABCS~\cite{ABCS} experiment 
(Fig.~\ref{fig:ccorxt}d) are in excellent agreement due 
to the cancellation of the systematic errors in each experiment. Thus, while the individual ISR experiments each provide a data set with common systematic uncertainties which cancel in scaling studies, there is no unique $\pi^0$ absolute cross section measurement from the ISR at 62.4 GeV which can be used as a comparison spectrum for RHIC measurements.

\section{Status of theory and experiment, circa 1982}
 
Hard-scattering was visible both at ISR and FNAL (Fixed Target) energies 
via inclusive single particle production at large $p_T\geq$ 2-3 
GeV/c. Scaling and dimensional arguments for plotting 
data revealed the systematics and underlying physics. The theorists had the 
basic underlying physics correct; but many (inconvenient) details remained to 
be worked out, several by experiment. The transverse momentum 
imbalance of outgoing parton-pairs, the ``$k_T$ effect", was 
discovered by experiment~\cite{CCHK,MJT79}. The first modern QCD calculation and 
prediction for high $p_T$ single particle inclusive cross sections, including 
non-scaling and initial state radiation was done in 1978, by Jeff 
Owens and collaborators~\cite{Owens78} under the assumption that high $p_T$ particles  
are produced from states with two roughly back-to-back jets
which are the result of scattering of constituents of the nucleons (partons). 
   The overall $p+p$ hard-scattering cross section in ``leading logarithm" pQCD   
is the sum over parton reactions $a+b\rightarrow c +d$ 
(e.g. $g+q\rightarrow g+q$) at parton-parton center-of-mass (c.m.) energy $\sqrt{\hat{s}}=\sqrt{x_1 x_2 s}$.  
\begin{equation}
\frac{d^3\sigma}{dx_1 dx_2 d\cos\theta^*}=
\frac{1}{s}\sum_{ab} f_a(x_1) f_b(x_2) 
\frac{\pi\alpha_s^2(Q^2)}{2x_1 x_2} \Sigma^{ab}(\cos\theta^*)
\label{eq:QCDabscat}
\end{equation} 
where $f_a(x_1)$, $f_b(x_2)$, are parton distribution functions, 
the differential probabilities for partons
$a$ and $b$ to carry momentum fractions $x_1$ and $x_2$ of their respective 
protons (e.g. $u(x_2)$), and where $\theta^*$ is the scattering angle in the parton-parton c.m. system. The characteristic subprocess angular distributions,
{\bf $\Sigma^{ab}(\cos\theta^*)$},
and the coupling constant,
$\alpha_s(Q^2)=\frac{12\pi}{25} \ln(Q^2/\Lambda^2)$,
are fundamental predictions of QCD~\cite{CutlerSivers,Combridge:1977dm}.

\begin{figure}[ht]
\begin{center}
\begin{tabular}{cc}
\hspace*{-0.1in}\includegraphics[width=0.75\linewidth]{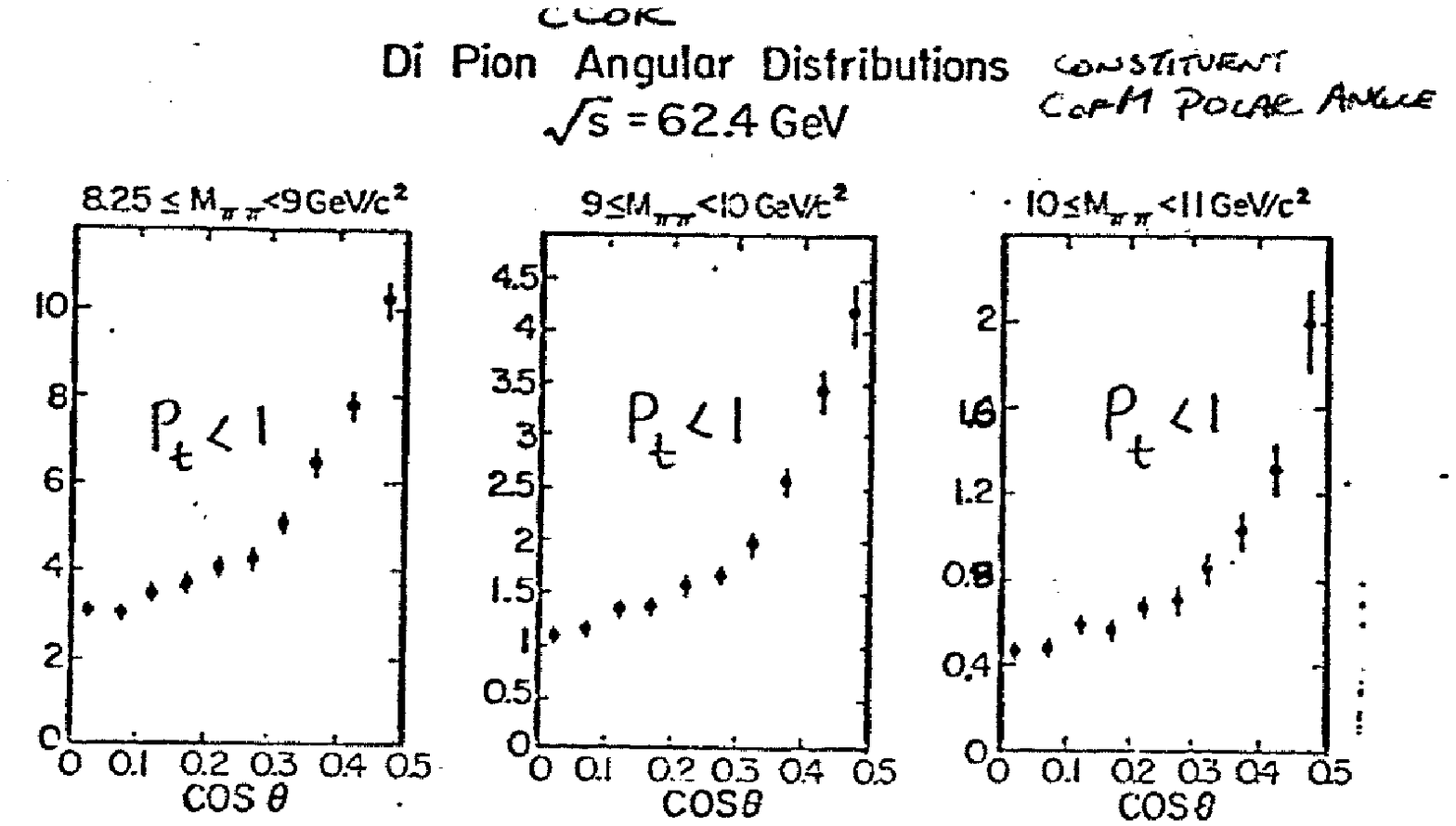} &
\hspace*{-0.35in}\includegraphics[width=0.288\linewidth]{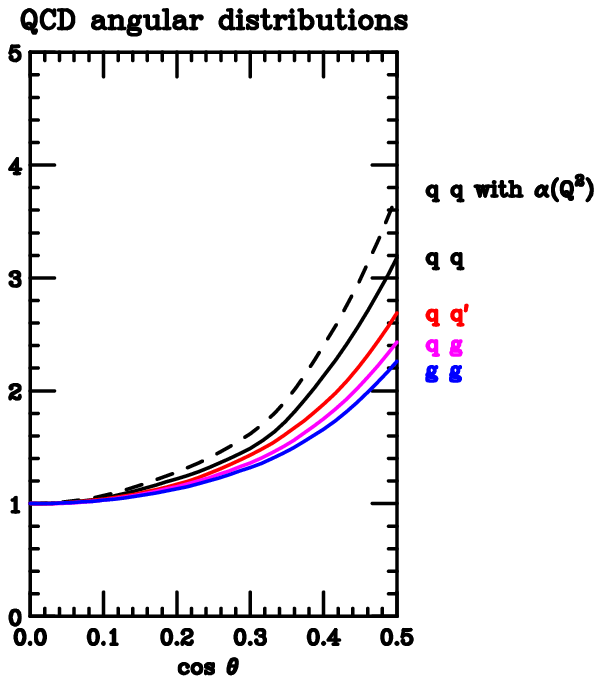}
\end{tabular}
\end{center}
\vspace*{-10mm}
\caption[]
{a) (left 3 panels) CCOR measurement~\cite{Paris82,CCOR82NPB} of polar angular distributions of $\pi^0$ pairs with net $p_T < 1$ GeV/c at mid-rapidity in p-p collisions with $\sqrt{s}=62.4$ GeV for 3 different values of $\pi\pi$ invariant mass $M_{\pi \pi}$. b) (rightmost panel) QCD predictions for $\Sigma^{ab}(\cos\theta^*)$ for the elastic scattering of $gg$, $qg$, $qq'$, $qq$, and $qq$ with $\alpha_s(Q^2)$ evolution.    
\label{fig:ccorqq} }
\end{figure}

 However, jets in $4\pi$ calorimeters at ISR 
energies or lower are invisible below $\sqrt{\hat{s}}\sim E_T \leq 25$ 
GeV~\cite{Gordon}. Nevertheless, there were many false claims of jet observation in the period 1977-1982 which led to skepticism 
about jets in hadron collisions, particularly in the USA~\cite{MJTIJMPA}. 
A `phase change' in belief-in-jets was produced by one UA2 event 
at the 1982 ICHEP in Paris~\cite{Paris82}, which, together with the first direct measurement of the QCD constituent-scattering angular distribution, $\Sigma^{ab}(\cos\theta^*)$ (Eq.~\ref{eq:QCDabscat}), using two-particle correlations, presented at the same meeting (Fig.~\ref{fig:ccorqq}), gave universal credibility to the pQCD description of high $p_T$ hadron physics~\cite{Owens,Darriulat,DiLella}.    

\section{Almost everything you want to know about jets can be found using 2-particle correlations.} 

   The outgoing jet-pairs of hard-scattering obey the kinematics of elastic-scattering (of partons) in a parton-parton c.m. frame which is longitudinally moving with rapidity $y=1/2 \ln(x_1/x_2)$ in the p-p c.m. frame. Hence, the jet pair formed from the scattered partons should be co-planar with the beam axis, with equal and opposite transverse momenta, and thus be back-to-back in azimuthal projection. It is not necessary to fully reconstruct the jets in order to measure their properties. In many cases two-particle correlations are sufficient to measure the desired properties, and in some cases, such as the measurement of the net $p_T$ of a jet-pair, may be superior, since the issue of the systematic error caused by missing some of the particles in the jet is not-relevant. A helpful property in this regard is the ``leading-particle effect''. Due to the steeply falling power-law transverse momentum spectrum of the scattered partons, the inclusive single particle (e.g. $\pi$) spectrum from jet fragmentation is dominated by fragments with large $z$, where $z=p_{T\pi}/p_{T_q}$ is the fragmentation variable. The probability for a fragment pion, with momentum fraction $z$, from a parton with $p_{T_q}=p_{T{\rm jet}}$ is:
   \begin{equation}
   {{d^2\sigma_{\pi} (p_{T_q},z) }\over {dp_{T_q} dz }}={{d\sigma_q}\over {dp_{T_q}}}\times D^q_{\pi} (z)={A \over {p_{T_q}^{m-1}}}  \times D^q_{\pi} (z) 
 \qquad ,  \label{eq:zgivenq}
   \end{equation}
   where $D^q_{\pi} (z)\sim e^{-6z}$ is the fragmentation function. 
The change of variables, $p_{T_q}=p_{T_{\pi}}/z$, ${dp_{T_q}}/{dp_{T_{\pi}}}|_{z}=1/z$, then gives the joint probability of a fragment $\pi$, with  transverse momentum $p_{T_{\pi}}$ and fragmentation fraction $z$: 
\begin{equation}
{{d^2\sigma_{\pi} (p_{T_{\pi}},z)} \over { dp_{T_{\pi}} dz}} 
={A \over {p_{T_{\pi}}^{m-1}}} \times z^{m-2} D^q_{\pi} (z) \qquad . 
\label{eq:zgivenpi}
\end{equation}
Thus, the effective fragmentation function, given that a fragment (with $p_{T_{\pi}}$) is detected, is weighted upward in $z$ by a factor $z^{m-2}$, where $m$ is the simple power fall-off of the jet invariant cross section (i.e. not the $n(x_T, \sqrt{s})$ of Eq.~\ref{eq:bbg}~\cite{confusion}). As this property, although general, is most useful in studying `unbiased' away jets using biased trigger jets selected by single particle triggers, it was given the unfortunate name `trigger-bias'~\cite{JacobLandshoff}. 

\begin{figure}[ht]
\begin{center}
\includegraphics[width=0.50\linewidth]{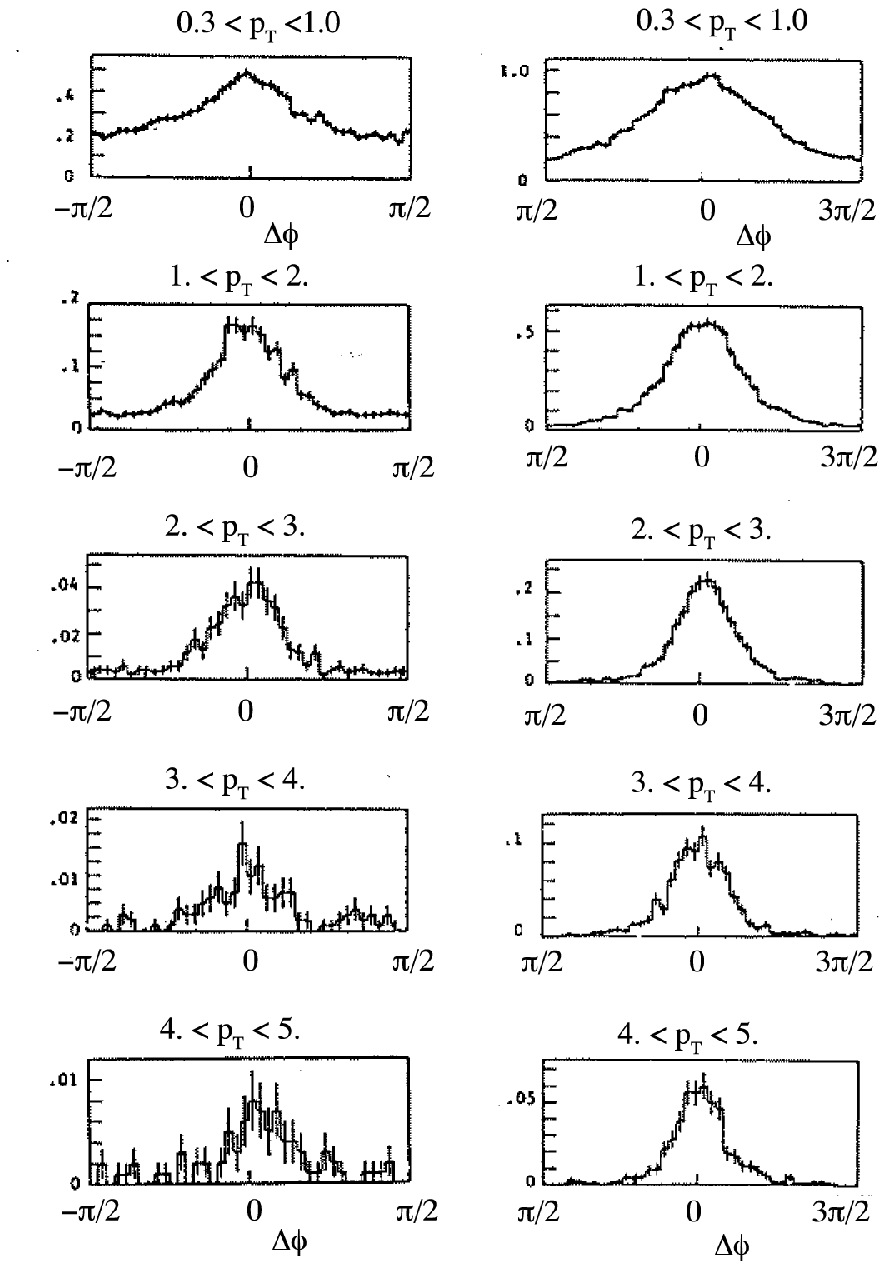} 
\includegraphics[width=0.48\linewidth]{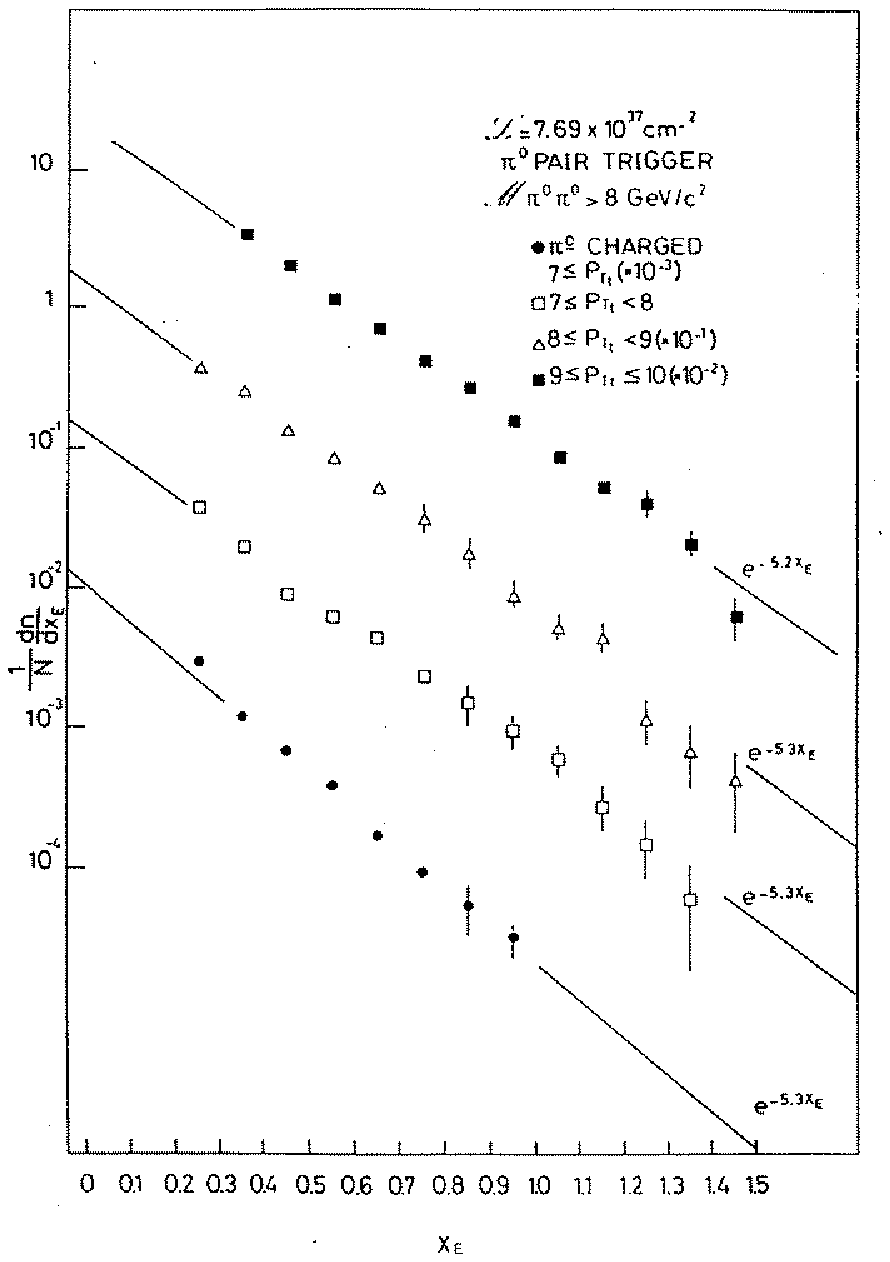}
\end{center}
\vspace*{-10mm}
\caption[]
{a,b) Azimuthal distributions of charged particles of transverse momentum $p_T$, with respect to a trigger $\pi^0$ with $p_{Tt}\geq 7$ GeV/c, for 5 intervals of $p_T$: a) (left-most panel) for $\Delta\phi=\pm \pi/2$ rad about the trigger particle, and b) (middle panel) for $\Delta\phi=\pm \pi/2$ about $\pi$ radians (i.e. directly opposite in azimuth) to the trigger. The trigger particle is restricted to $|\eta|<0.4$, while the associated charged particles are in the range $|\eta|\leq 0.7$. c) (right panel) $x_E$ distributions (see text) corresponding to the data of the center panel.   
\label{fig:ccorazi} }
\end{figure}

\begin{figure}[ht]
\begin{center}
\begin{tabular}{cc}
\includegraphics[width=0.47\linewidth]{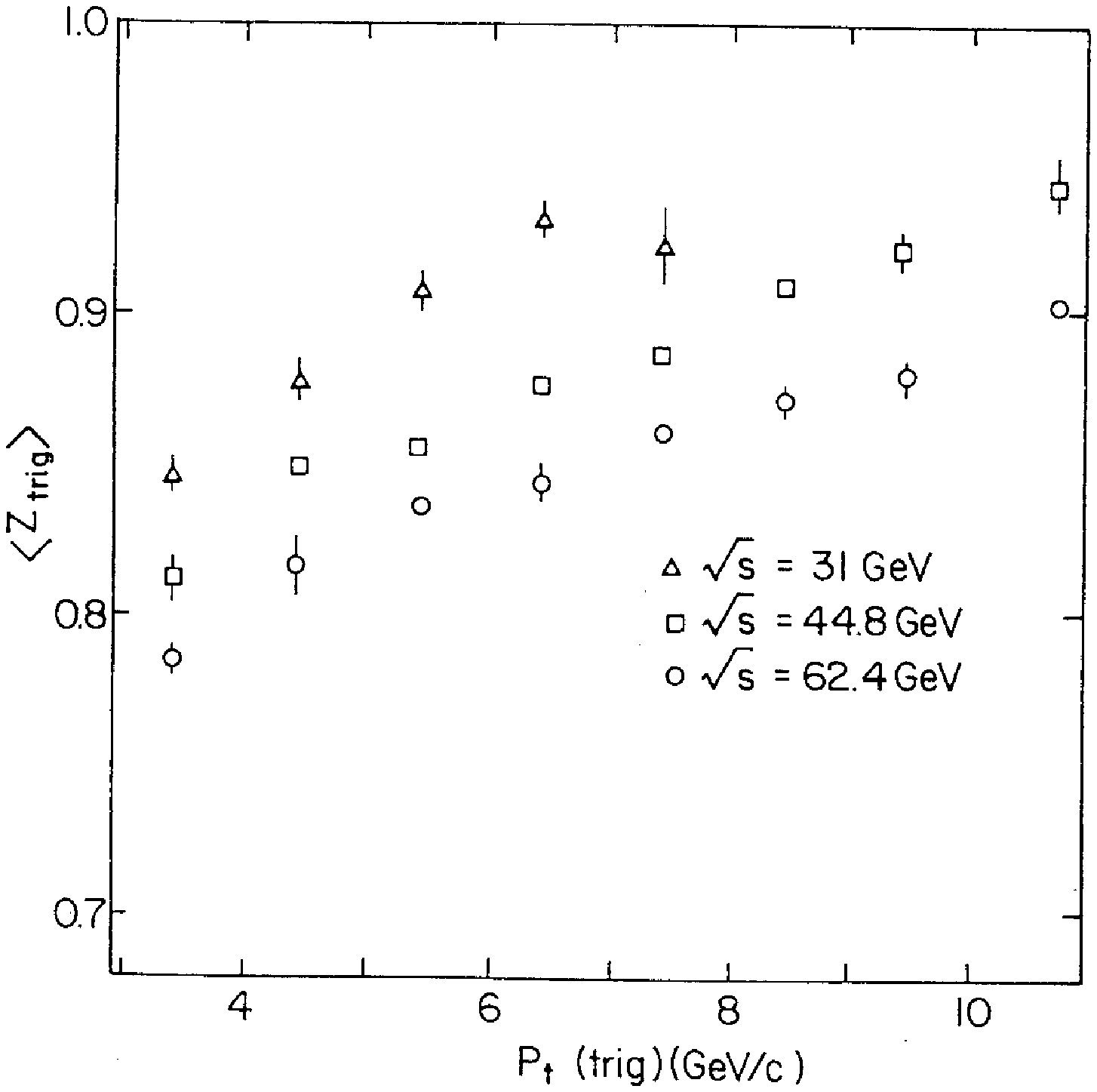} &
\includegraphics[width=0.50\linewidth,angle=-1]{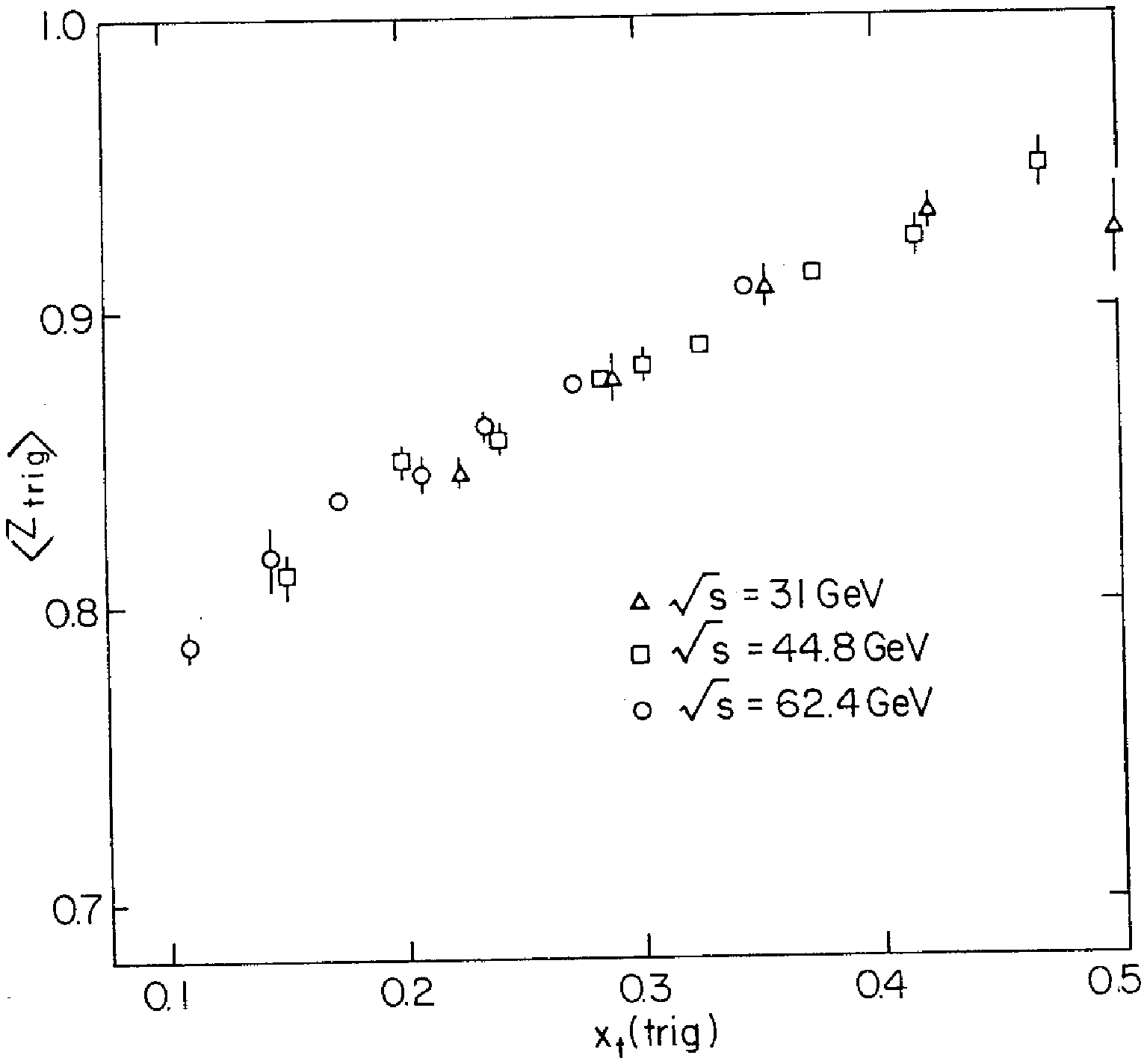}
\end{tabular}
\end{center}
\vspace*{-10mm}
\caption[]
{CCOR~\cite{CCOR82NPB} measurement of $\langle z_{\rm trig}\rangle$ as a function of $p_{Tt}$ (left) and $x_{Tt}=2p_{Tt}/\sqrt{s}$ (right).  
\label{fig:ccormeanz} }
\end{figure}

   Many ISR experiments provided excellent 2-particle correlation measurements~\cite{Moriond79}. However, the CCOR experiment~\cite{Angelis79} was the first to provide charged particle measurement with full and uniform acceptance over the entire azimuth, with pseudorapidity coverage $-0.7\leq \eta\leq 0.7$, so that the jet structure of high $p_T$ scattering could be easily seen and measured. In  Fig.~\ref{fig:ccorazi}a,b, the azimuthal distributions of associated charged particles 
relative to a $\pi^0$ trigger with transverse momentum $p_{Tt} > 7$ GeV/c are shown for five intervals of associated particle transverse momentum $p_T$. In all cases, strong correlation peaks on flat backgrounds are clearly visible, indicating the di-jet structure which is contained in an interval $\Delta\phi=\pm 60^\circ$ about a direction towards and opposite the to trigger for all values of associated $p_T\, (>0.3$ GeV/c) shown. The width of the peaks about the trigger direction (Fig.~\ref{fig:ccorazi}a), or opposite to the trigger (Fig.~\ref{fig:ccorazi}b) indicates out-of-plane activity from the individual fragments of jets. The trigger bias was directly measured from these data by reconstructing the trigger jet from the associated charged particles with $p_T\geq 0.3$ Gev/c, within $\Delta\phi=\pm 60^\circ$ from the trigger particle, using the algorithm $p_{T{\rm jet}}=p_{Tt}+1.5\sum p_T\cos(\Delta\phi)$, where the factor 1.5 corrects the measured charged particles for missing neutrals. The measured $z_{\rm trig}=p_{Tt}/p_{T{\rm jet}}$ distributions for 3 values of $\sqrt{s}$ (Fig.~\ref{fig:ccormeanz}) show the ``unexpected''~\cite{JacobEPS79} property of $x_T$ scaling.  

\begin{figure}[ht]
\begin{center}
\begin{tabular}{cc}
\includegraphics[width=0.47\linewidth]{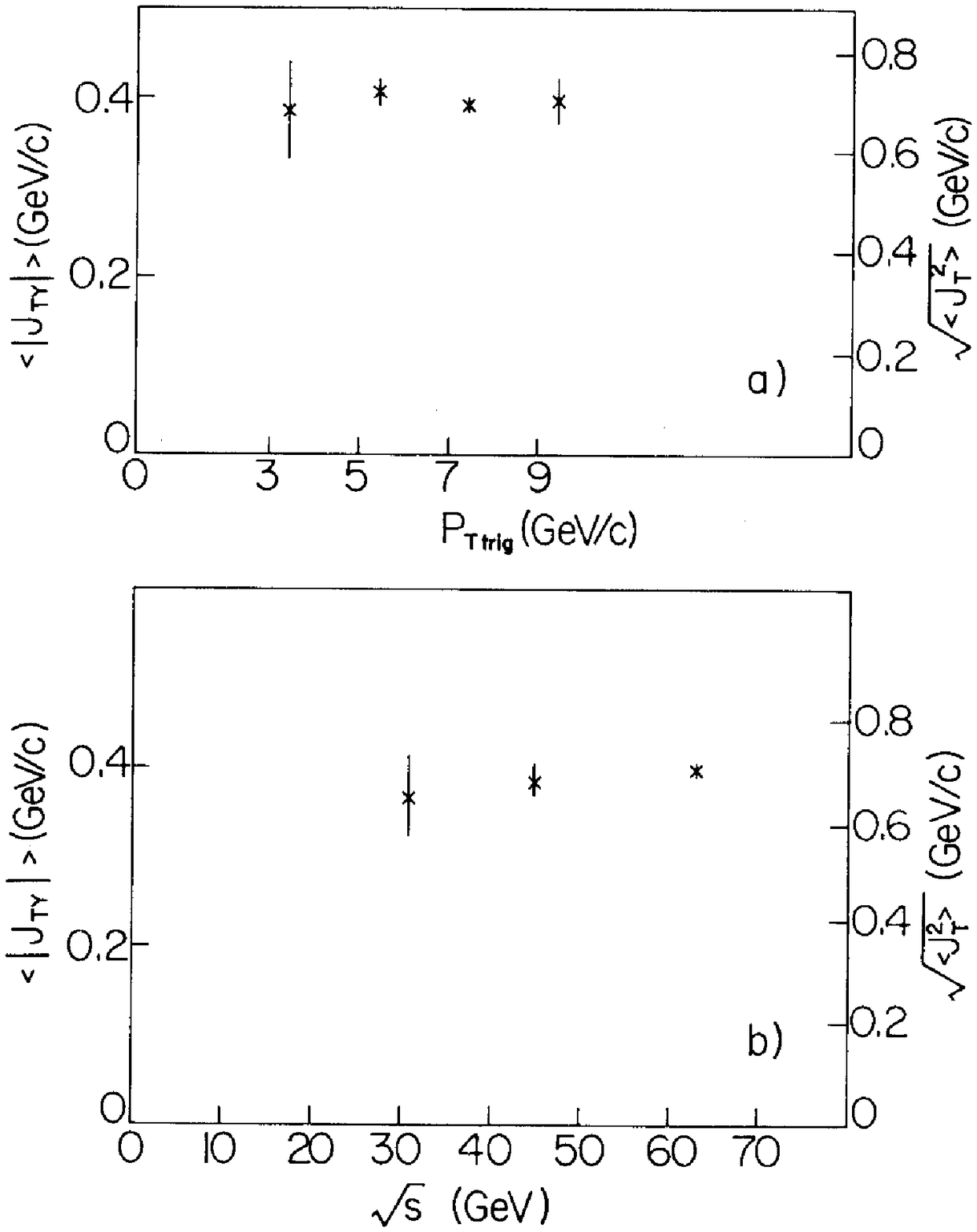} &
\includegraphics[width=0.50\linewidth]{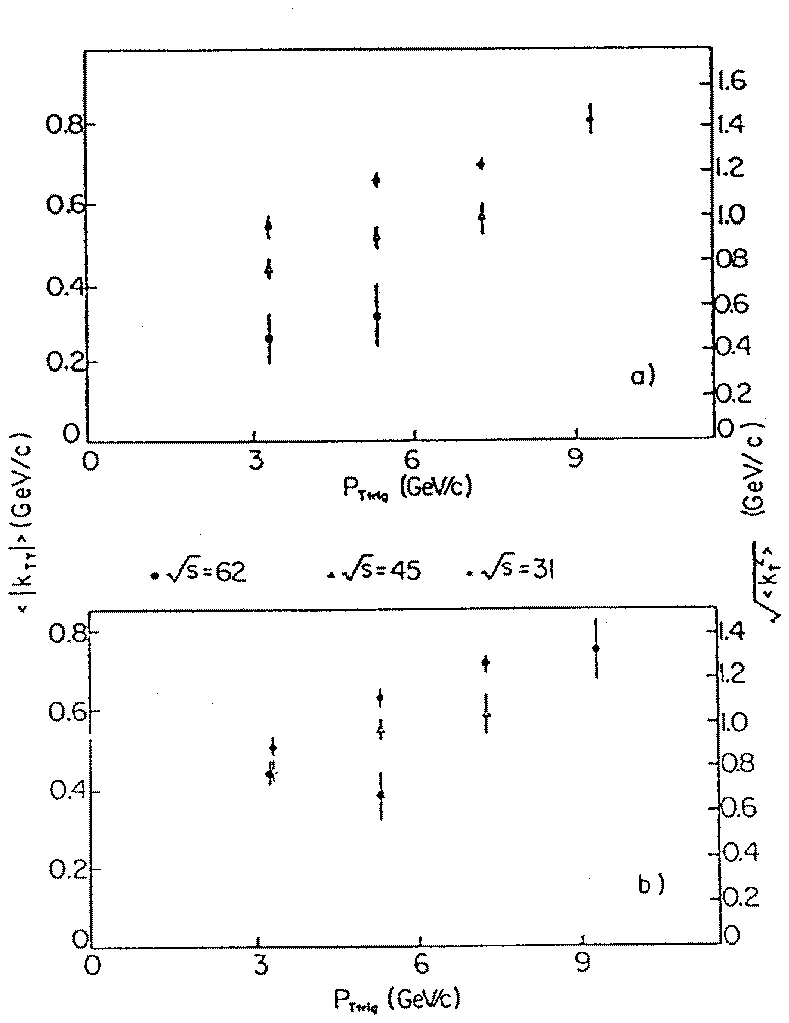}
\end{tabular}
\end{center}
\vspace*{-10mm}
\caption[]
{CCOR~\cite{CCOR80} measurements of $\langle j_{T_y}\rangle$ (left) $\langle k_{T_y}\rangle$ (right) as a function of $p_{Tt}$ for 3 values of $\sqrt{s}$.   
The mean absolute values of the components $j_{T_y}\equiv j_{T_{\phi}}$ are related to $\sqrt{\langle j_{T}^2}\rangle$ with the assumption that $j_{T_x}$ and $j_{T_y}$ are independent gaussians, with equal r.m.s., which combine independently to form $j_T^2=j_{T_x}^2 + j_{T_y}^2$~\cite{Ap99}.    
\label{fig:ccorjtkt} }
\end{figure}

  	Following the analysis of previous CERN-ISR experiments~\cite{Darriulat76,CCHK}, the away jet azimuthal angular distributions  of Fig.~\ref{fig:ccorazi}b, which should be unbiased, were analyzed in terms of the two variables: $p_{\rm out}=p_T \sin(\Delta\phi)$, the out-of-plane transverse momentum of a track, 
 and $x_E$, where:\\ 
\begin{minipage}[c]{0.5\linewidth}
\vspace*{-0.36in}
\begin{equation}	
x_E=\frac{-\vec{p}_T\cdot \vec{p}_{Tt}}{|p_{Tt}|^2}=\frac{-p_T \cos(\Delta\phi)}{p_{Tt}}\simeq \frac {z}{z_{\rm trig}}  
\label{eq:xE}
\end{equation}
\end{minipage}
\hspace*{0.01\linewidth}
\begin{minipage}[b]{0.50\linewidth}
\includegraphics[scale=0.6]{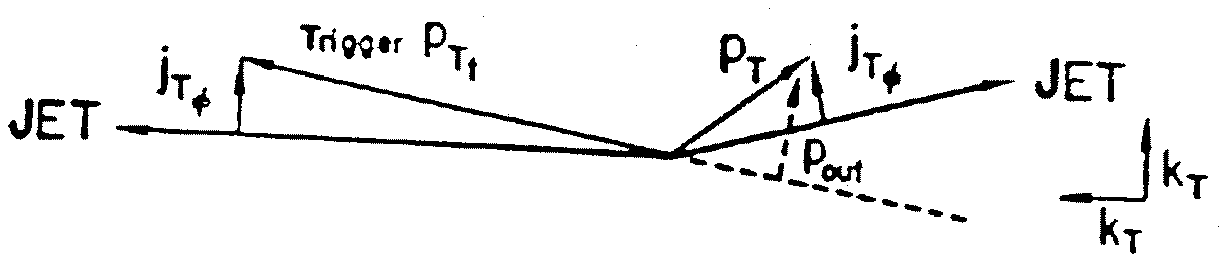}
\vspace*{-0.12in}
\label{fig:poutxe}
\end{minipage}
$z_{\rm trig}\simeq p_{Tt}/p_{T{\rm jet}}$ is the fragmentation variable of the trigger jet, and $z$ is the fragmentation variable of the away jet. Note that $x_E$ would equal the fragmenation fraction $z$ of the away jet, for $z_{\rm trig}\rightarrow 1$, if the trigger and away jets balanced transverse momentum. 
The $x_E$ distributions~\cite{Angelis79,JacobEPS79b} for the data of Fig.~\ref{fig:ccorazi}b are shown in Fig.~\ref{fig:ccorazi}c and show the expected fragmentation behavior, $e^{-6z}\sim e^{-6 x_E \langle z_{\rm trig}\rangle}$. If the width of the away distributions (Fig.~\ref{fig:ccorazi}b) corresponding to the out of plane activity were due entirely to jet fragmentation, then  
$\langle |\sin(\Delta\phi)|\rangle=\langle |j_{T_{\phi}}|/p_T \rangle$ would decrease in direct proportion to $1/p_T$, where $j_{T_{\phi}}$ is the component of $\vec{j}_T$ in the azimuthal plane, since the jet fragmentation transverse momentum, $\vec{j}_T$, should be independent of $p_T$.  This is clearly not the case, as originally shown by the CCHK collaboration~\cite{CCHK}, which inspired Feynman, Field and Fox (FFF)~\cite{FFF} to introduce, $\vec{k}_T$, the transverse momentum of a parton in a nucleon. In this formulation, the net transverse momentum of an outgoing parton pair is $\sqrt{2} k_T$, which is composed of two orthogonal components, $\sqrt{2} k_{T_{\phi}}$, out of the scattering plane, which makes the jets acoplanar, i.e. not back-to-back in azimuth, and $\sqrt{2} k_{T_x}$, along the axis of the trigger jet, which makes the jets unequal in energy. Originally, ${k}_T$ was thought of as having an `intrinsic' part from confinement, which would be constant as a function of $x$ and $Q^2$, and a part from NLO hard-gluon emission, which would vary with $x$ and $Q^2$, however now it is explained as `resummation' to all orders of QCD~\cite{Sterman}. 
	FFF~\cite{FFF,Levin} gave the approximate formula to derive $k_T$ from the measurement of $p_{\rm out}$ as a function of $x_E$:
\begin{equation}
\langle |p_{\rm out}|\rangle^2=x_E^2 [2\langle |k_{T_{\phi}}|\rangle^2 +  \langle |j_{T_{\phi}}|\rangle^2 ] + \langle |j_{T_{\phi}}|\rangle^2 \qquad .
\label{eq:FFFpoutkT}
\end{equation}
CCOR~\cite{CCOR80} used this formula to derive $\langle |k_{T_{\phi}}|\rangle$ and $\langle |j_{T_{\phi}}|\rangle$ as a function of $p_{Tt}$ and $\sqrt{s}$ from the data of Fig.~\ref{fig:ccorazi}b (see Fig.~\ref{fig:ccorjtkt}). This important result shows that $\langle |j_{T_{\phi}}|\rangle$ is constant, independent of $p_{Tt}$ and $\sqrt{s}$, as expected for fragmentation, but that $\langle |k_{T_{\phi}}|\rangle$ varies with both $p_{Tt}$ and $\sqrt{s}$, suggestive of a radiative, rather than an intrinsic origin for $k_T$. 
\section{Conclusion} 
   It should be noted that inclusion of $k_T$ was the key element~\cite{Owens78}  beyond QCD to explain the $n\simeq8$ $x_T$-scaling result of the CCR~\cite{CCR} and FNAL (fixed target) experiments~\cite{Cronin}. More recent FNAL fixed target measurements~\cite{Ap99} and many theoretical works have used $k_T$ as an empirical parameter to improve the comparison of measurement to NLO QCD. However, it is important to remember, as illustrated above, that $k_T$ is not simply a parameter, it can be measured.

\begin{thebibliography}{99}
\bibitem{confusion} However, this delicacy is often glibly ignored.
\end{thebibliography}
\end{document}